\documentclass[aps,prl,twocolumn,superscriptaddress,showpacs]{revtex4} % UTOBS

\usepackage{graphicx}
\usepackage{rotating}

\begin{document}

%\pagewiselinenumbers  % LINENO CTOBS

%\newcommand{\fix}[1]{{\bf{} }{\{\it #1}\}} %  comment this out if you don't want to see the comments
\newcommand{\fix}[1]{} % comment this in if you want to the see the PRL without comments
\newcommand{\MET}{\mbox{$\raisebox{.3ex}{$\not\!$}E_T$}}
\newcommand{\METVEC}{\mbox{$\raisebox{.3ex}{$\not\!$}{\vec E}_T$}}
\newcommand{\MPTVEC}{\mbox{$\raisebox{.3ex}{$\not\!$}{\vec {p}_T}$}}

\def\DZero{D0~}

%%%% MET +JETS %$$$$$$$$$
\newcommand{\Et}{\mbox{$E_T$}}
\newcommand{\Pt}{\mbox{$p_T$}}
\newcommand{\Ht}{\mbox{$H_T$}}
\newcommand{\met}{\mbox{$\protect \raisebox{0.3ex}{$\not$}\Et$}}
\newcommand{\mpt}{\mbox{$\protect \raisebox{0.3ex}{$\not$}\Pt$}}

%\preprint{CDF/DOC/TOP/CDFR/9361} %Comment this out before shipping out
%\preprint{Version 2.1}          %Comment this out before shipping out

% Question marks indicate numbers to be filled in -- there should be no question marks
% in the final draft.

\title{Measurement of the WW+WZ Production Cross Section Using the Lepton + Jets Final State at CDF II}

%%This gets replaced by authors and institutions
%\vspace{0.5in}
%\author{The CDF Collaboration} % This will get replaced with the CDF author list. 
                               % See the GP webpage to get it.

\affiliation{Institute of Physics, Academia Sinica, Taipei, Taiwan 11529, Republic of China} 
\affiliation{Argonne National Laboratory, Argonne, Illinois 60439} 
\affiliation{University of Athens, 157 71 Athens, Greece} 
\affiliation{Institut de Fisica d'Altes Energies, Universitat Autonoma de Barcelona, E-08193, Bellaterra (Barcelona), Spain} 
\affiliation{Baylor University, Waco, Texas  76798} 
\affiliation{Istituto Nazionale di Fisica Nucleare Bologna, $^{cc}$University of Bologna, I-40127 Bologna, Italy} 
\affiliation{Brandeis University, Waltham, Massachusetts 02254} 
\affiliation{University of California, Davis, Davis, California  95616} 
\affiliation{University of California, Los Angeles, Los Angeles, California  90024} 
\affiliation{University of California, San Diego, La Jolla, California  92093} 
\affiliation{University of California, Santa Barbara, Santa Barbara, California 93106} 
\affiliation{Instituto de Fisica de Cantabria, CSIC-University of Cantabria, 39005 Santander, Spain} 
\affiliation{Carnegie Mellon University, Pittsburgh, PA  15213} 
\affiliation{Enrico Fermi Institute, University of Chicago, Chicago, Illinois 60637}
\affiliation{Comenius University, 842 48 Bratislava, Slovakia; Institute of Experimental Physics, 040 01 Kosice, Slovakia} 
\affiliation{Joint Institute for Nuclear Research, RU-141980 Dubna, Russia} 
\affiliation{Duke University, Durham, North Carolina  27708} 
\affiliation{Fermi National Accelerator Laboratory, Batavia, Illinois 60510} 
\affiliation{University of Florida, Gainesville, Florida  32611} 
\affiliation{Laboratori Nazionali di Frascati, Istituto Nazionale di Fisica Nucleare, I-00044 Frascati, Italy} 
\affiliation{University of Geneva, CH-1211 Geneva 4, Switzerland} 
\affiliation{Glasgow University, Glasgow G12 8QQ, United Kingdom} 
\affiliation{Harvard University, Cambridge, Massachusetts 02138} 
\affiliation{Division of High Energy Physics, Department of Physics, University of Helsinki and Helsinki Institute of Physics, FIN-00014, Helsinki, Finland} 
\affiliation{University of Illinois, Urbana, Illinois 61801} 
\affiliation{The Johns Hopkins University, Baltimore, Maryland 21218} 
\affiliation{Institut f\"{u}r Experimentelle Kernphysik, Karlsruhe Institute of Technology, D-76131 Karlsruhe, Germany} 
\affiliation{Center for High Energy Physics: Kyungpook National University, Daegu 702-701, Korea; Seoul National University, Seoul 151-742, Korea; Sungkyunkwan University, Suwon 440-746, Korea; Korea Institute of Science and Technology Information, Daejeon 305-806, Korea; Chonnam National University, Gwangju 500-757, Korea; Chonbuk National University, Jeonju 561-756, Korea} 
\affiliation{Ernest Orlando Lawrence Berkeley National Laboratory, Berkeley, California 94720} 
\affiliation{University of Liverpool, Liverpool L69 7ZE, United Kingdom} 
\affiliation{University College London, London WC1E 6BT, United Kingdom} 
\affiliation{Centro de Investigaciones Energeticas Medioambientales y Tecnologicas, E-28040 Madrid, Spain} 
\affiliation{Massachusetts Institute of Technology, Cambridge, Massachusetts  02139} 
\affiliation{Institute of Particle Physics: McGill University, Montr\'{e}al, Qu\'{e}bec, Canada H3A~2T8; Simon Fraser University, Burnaby, British Columbia, Canada V5A~1S6; University of Toronto, Toronto, Ontario, Canada M5S~1A7; and TRIUMF, Vancouver, British Columbia, Canada V6T~2A3} 
\affiliation{University of Michigan, Ann Arbor, Michigan 48109} 
\affiliation{Michigan State University, East Lansing, Michigan  48824}
\affiliation{Institution for Theoretical and Experimental Physics, ITEP, Moscow 117259, Russia} 
\affiliation{University of New Mexico, Albuquerque, New Mexico 87131} 
\affiliation{Northwestern University, Evanston, Illinois  60208} 
\affiliation{The Ohio State University, Columbus, Ohio  43210} 
\affiliation{Okayama University, Okayama 700-8530, Japan} 
\affiliation{Osaka City University, Osaka 588, Japan} 
\affiliation{University of Oxford, Oxford OX1 3RH, United Kingdom} 
\affiliation{Istituto Nazionale di Fisica Nucleare, Sezione di Padova-Trento, $^{dd}$University of Padova, I-35131 Padova, Italy} 
\affiliation{LPNHE, Universite Pierre et Marie Curie/IN2P3-CNRS, UMR7585, Paris, F-75252 France} 
\affiliation{University of Pennsylvania, Philadelphia, Pennsylvania 19104}
\affiliation{Istituto Nazionale di Fisica Nucleare Pisa, $^{ee}$University of Pisa, $^{ff}$University of Siena and $^{gg}$Scuola Normale Superiore, I-56127 Pisa, Italy} 
\affiliation{University of Pittsburgh, Pittsburgh, Pennsylvania 15260} 
\affiliation{Purdue University, West Lafayette, Indiana 47907} 
\affiliation{University of Rochester, Rochester, New York 14627} 
\affiliation{The Rockefeller University, New York, New York 10021} 
\affiliation{Istituto Nazionale di Fisica Nucleare, Sezione di Roma 1, $^{hh}$Sapienza Universit\`{a} di Roma, I-00185 Roma, Italy} 

\affiliation{Rutgers University, Piscataway, New Jersey 08855} 
\affiliation{Texas A\&M University, College Station, Texas 77843} 
\affiliation{Istituto Nazionale di Fisica Nucleare Trieste/Udine, I-34100 Trieste, $^{ii}$University of Trieste/Udine, I-33100 Udine, Italy} 
\affiliation{University of Tsukuba, Tsukuba, Ibaraki 305, Japan} 
\affiliation{Tufts University, Medford, Massachusetts 02155} 
\affiliation{Waseda University, Tokyo 169, Japan} 
\affiliation{Wayne State University, Detroit, Michigan  48201} 
\affiliation{University of Wisconsin, Madison, Wisconsin 53706} 
\affiliation{Yale University, New Haven, Connecticut 06520} 
\author{T.~Aaltonen}
\affiliation{Division of High Energy Physics, Department of Physics, University of Helsinki and Helsinki Institute of Physics, FIN-00014, Helsinki, Finland}
\author{J.~Adelman}
\affiliation{Enrico Fermi Institute, University of Chicago, Chicago, Illinois 60637}
\author{B.~\'{A}lvarez~Gonz\'{a}lez$^v$}
\affiliation{Instituto de Fisica de Cantabria, CSIC-University of Cantabria, 39005 Santander, Spain}
\author{S.~Amerio$^{dd}$}
\affiliation{Istituto Nazionale di Fisica Nucleare, Sezione di Padova-Trento, $^{dd}$University of Padova, I-35131 Padova, Italy} 

\author{D.~Amidei}
\affiliation{University of Michigan, Ann Arbor, Michigan 48109}
\author{A.~Anastassov}
\affiliation{Northwestern University, Evanston, Illinois  60208}
\author{A.~Annovi}
\affiliation{Laboratori Nazionali di Frascati, Istituto Nazionale di Fisica Nucleare, I-00044 Frascati, Italy}
\author{J.~Antos}
\affiliation{Comenius University, 842 48 Bratislava, Slovakia; Institute of Experimental Physics, 040 01 Kosice, Slovakia}
\author{G.~Apollinari}
\affiliation{Fermi National Accelerator Laboratory, Batavia, Illinois 60510}
\author{A.~Apresyan}
\affiliation{Purdue University, West Lafayette, Indiana 47907}
% ...

\collaboration{CDF Collaboration\footnote{With visitors from $^a$University of Massachusetts Amherst, Amherst, Massachusetts 01003,
$^b$Universiteit Antwerpen, B-2610 Antwerp, Belgium, 
$^c$University of Bristol, Bristol BS8 1TL, United Kingdom,
$^d$Chinese Academy of Sciences, Beijing 100864, China, 
$^e$Istituto Nazionale di Fisica Nucleare, Sezione di Cagliari, 09042 Monserrato (Cagliari), Italy,
$^f$University of California Irvine, Irvine, CA  92697, 
$^g$University of California Santa Cruz, Santa Cruz, CA  95064, 
$^h$Cornell University, Ithaca, NY  14853, 
$^i$University of Cyprus, Nicosia CY-1678, Cyprus, 
$^j$University College Dublin, Dublin 4, Ireland,
$^k$University of Edinburgh, Edinburgh EH9 3JZ, United Kingdom, 
$^l$University of Fukui, Fukui City, Fukui Prefecture, Japan 910-0017
$^m$Kinki University, Higashi-Osaka City, Japan 577-8502
$^n$Universidad Iberoamericana, Mexico D.F., Mexico,
$^o$University of Iowa, Iowa City, IA  52242,
$^p$Kansas State University, Manhattan, KS 66506
$^q$Queen Mary, University of London, London, E1 4NS, England,
$^r$University of Manchester, Manchester M13 9PL, England,
$^s$Muons, Inc., Batavia, IL 60510, 
$^t$Nagasaki Institute of Applied Science, Nagasaki, Japan, 
$^u$University of Notre Dame, Notre Dame, IN 46556,
$^v$University de Oviedo, E-33007 Oviedo, Spain, 
$^w$Texas Tech University, Lubbock, TX  79609, 
$^x$IFIC(CSIC-Universitat de Valencia), 56071 Valencia, Spain,
$^y$Universidad Tecnica Federico Santa Maria, 110v Valparaiso, Chile,
$^z$University of Virginia, Charlottesville, VA  22906
$^{aa}$Bergische Universit\"at Wuppertal, 42097 Wuppertal, Germany,
$^{bb}$Yarmouk University, Irbid 211-63, Jordan
$^{jj}$On leave from J.~Stefan Institute, Ljubljana, Slovenia, 
}}
\noaffiliation

%\date{March 4, 2009}
%\today

\begin{abstract}
We report two complementary measurements of the $WW+WZ$ cross section
in the final state consisting of an electron or muon, missing transverse energy, and jets,
performed using $p\bar{p}$ collision data at $\sqrt{s} = $1.96~TeV collected by the CDF II detector.
The first method uses the dijet invariant mass distribution while
%3.9 fb$^{-1}$ of integrated luminosity
%The resulting $WW+WZ$ production cross section measurement is
%$\sigma_{WW+WZ} = 14.4 \pm 3.1{\rm (stat)}\pm 2.2 {\rm (syst)}\ {\rm pb}$.
the second more sensitive method uses %more of the kinematic information in the
%event through 
matrix-element calculations.  %of the signal and background
%processes 
%and has a higher sensitivity.
%Using 2.7 fb$^{-1}$ of integrated luminosity the resulting
%measurement is $\sigma_{WW+WZ} = 17.7 \pm 3.1{\rm (stat)}\pm 2.4 {\rm (syst)}\ {\rm pb}$.
The result from the second method has a signal significance of 5.4$\sigma$ and is the 
first observation of $WW+WZ$ production using this signature.
Combining the results gives $\sigma_{WW+WZ} = 16.0 \pm 3.3\ {\rm pb}$,
in agreement with the standard model prediction.
%Both results
%are in good agreement with each other and with the standard model prediction.

\end{abstract}

  %% using the PACS 2008 list.
%%  SM Higgs boson
%% W bosons
%% Z bosons
%% Applications of electroweak models to specific processes

%\keywords{}

\maketitle


%\section{Introduction}

Measurements involving heavy vector boson pairs ($WW$, $WZ$, and $ZZ$) are
important tests of the electroweak sector of the standard model (SM).
Deviations of the production cross section from predictions could
arise from anomalous triple gauge boson interactions~\cite{hagiwara}
or from new resonances decaying to vector bosons.
Furthermore, the topology of diboson events is similar to that of events in which
a Higgs boson is produced in association with a $W$ or a $Z$,
allowing diboson measurements to provide an important step 
towards future measurements of Higgs boson production.

%Diboson production has been observed at the Tevatron in channels where
%both bosons decay leptonically~\cite{diblepCDF,diblepD0}.
%Extraction of the diboson signal in hadronic channels is more
%challenging because of significantly larger backgrounds. 
%%We assume the SM ratio of $WW$ and $WZ$ production in .... 
%The CDF collaboration has reported first observation of $WW + WZ + ZZ$
%production in a hadronic channel with large missing transverse energy
%and jets~\cite{cdfmetjj}. We report on two measurements of the cross
%section, $\sigma (p\bar{p} \rightarrow WW+WZ)$, with the CDF II
%detector that use different techniques in a channel with
%an identified lepton and jets.  The signal is the sum of $WW$ and $WZ$
%events because limited detector resolution makes it difficult to
%distinguish hadronically decaying $W$ bosons from $Z$ bosons. Evidence
%for $WW+WZ$ production in the lepton + jets decay mode has previously
%been reported by the D\O\ collaboration~\cite{d0lvjj} and the CDF
%collaboration recently set a limit on its cross section times
%branching ratio~\cite{cdflvjj}.  Our result represents the first
%observation of diboson production in this decay channel.

Diboson production has been observed at the Tevatron in channels in which
both bosons decay leptonically~\cite{diblepCDF,diblepD0}.
Extraction of the diboson signal in hadronic channels is more
challenging because of significantly larger backgrounds. In addition, due to 
limited detector resolution, it is difficult to distinguish 
hadronically decaying $W$ bosons from $Z$ bosons.
We report on two measurements of the cross section,
$\sigma (p\bar{p} \rightarrow WW+WZ)$, with the CDF II detector~\cite{CDFdet} that use different techniques
applied to the leptonic decay
of one $W$ and the hadronic decay of the associated $W$ or $Z$
($WW/WZ \rightarrow \ell\nu qq$, where $\ell$ represents a high-$\Pt$ electron
or muon).
Our result represents the first observation of this signal in the lepton + jets channel.  Evidence has previously been 
reported by the \DZero collaboration~\cite{d0lvjj}, and the CDF collaboration set a limit on its cross section times branching ratio~\cite{cdflvjj}.
In addition the CDF collaboration has reported observation of $WW + WZ + ZZ$ in a different hadronic channel with large missing transverse energy and jets~\cite{cdfmetjj}. 
%%% should also mention this process has already been observed by CDF
%%% is a different decay mode: qqvv.

%The two measurements use different techniques
%applied to the leptonic decay
%of one $W$ and the hadronic decay of the associated $W$ or $Z$
%($WW/WZ \rightarrow \ell\nu qq$, where $\ell$ represents an electron
%or muon).

%%% should also mention this process has already been observed by CDF
%%% is a different decay mode: qqvv.

The first method uses the invariant mass of the two-jet system ($M_{jj}$)
to extract a signal peak from data corresponding to 3.9 fb$^{-1}$ of $p{\bar{p}}$ collisions at $\sqrt{s}=1.96$~TeV.
The second method takes advantage of more kinematic information
in the event by constructing a discriminant based on calculations
of the differential cross sections of the signal and background 
processes.  This so-called matrix-element (ME) method has been employed in 
a search for a low-mass Higgs produced in association with a $W$ 
boson~\cite{WHME} and in a measurement of single top production~\cite{cdfsingletop}.
%%% also single top, ZZ, WZ observation -- include these references
It is expected to achieve greater discriminating
power and here uses data corresponding to an integrated luminosity of 2.7 fb$^{-1}$.

%\section{CDF detector}

%The aspects of the CDF II detector~\cite{CDFdet}
%relevant to these analyses are briefly described here. The tracking
%system is composed of silicon microstrip detectors and an open-cell drift
%chamber inside a 1.4 T solenoid.  Electromagnetic
%lead-scintillator and hadronic iron-scintillator sampling calorimeters
%segmented in a projective geometry surround the tracking detectors.  A central
%calorimeter covers a pseudorapidity range of $|\eta| < 1.1$ while plug calorimeters 
%extend the acceptance into the region $1.1<|\eta|<3.6$.  Outside of the
%calorimeters are muon detectors composed of scintillators and drift
%chambers.  Cherenkov counters around the beam pipe and in the plug
%calorimeters count the inelastic collisions per bunch crossing and
%provide the luminosity measurement.

%\section{Trigger and event selection}

Data samples common to both analyses use trigger
selections requiring a central electron (muon) with $\Et$ ($\Pt$) $>$18 GeV.
The ME method utilizes an additional sample derived from a trigger requiring two jets and large missing transverse energy (\MET)~\cite{coord}.

Offline we select events with electron (muon) candidates
with $\Et$ ($\Pt$) $>$~20 GeV, and with \MET, jet, and other
kinematic requirements chosen differently for the two methods.
%congruous to the different methods used. 
Jets are clustered using a fixed-cone 
algorithm with radius 
$\Delta R = \sqrt{(\Delta\eta)^2+(\Delta\phi)^2} = 0.4$ and
their energies are corrected for detector effects~\cite{jet_details}.
Cosmic ray and photon conversion candidates are identified and removed.

Further event selection requirements are made to reduce backgrounds and
the sensitivity to systematic uncertainties. 
%These requirements are not the same for the two methods.
% because the $M_{jj}$
% analysis selection is driven by the demand that the $M_{jj}$ distribution
% be smoothly falling, while the ME analysis selection is driven by the need 
% to adequately model the quantities used in the matrix element calculation.
In the $M_{jj}$ method, we require events to have $\MET >25$~GeV,
at least two jets with $\Et >15$~GeV and $|\eta| < 2.4$, and the 
dijet vector boson candidate to have $\Pt > 40$ GeV/$c$.  As a result of these selection criteria, the $M_{jj}$ distribution for background is smoothly falling in the region where the signal is expected to peak.
The invariant mass of the dijet vector boson candidate, $M_{jj}$, is evaluated from the two most
energetic jets. 
Additional requirements
are made to reduce backgrounds and improve the Monte Carlo modeling of 
event kinematics:
the transverse mass of the lepton and \MET\ system ($M_{T}(W)$ ~\cite{coord}) must be greater
than 30 GeV/$c^2$, and the two most energetic jets must be separated
by $|\Delta \eta| < 2.5$.
%and the dijet vector boson candidate must have
%\Pt $>$~40 GeV/$c$.

%-------

In the ME method, we require events to have $\MET >$ 20~GeV
and exactly two jets with $\Et >25$~GeV and $|\eta| < 2.0$.  
Additional selection criteria to reduce backgrounds and 
achieve good modeling of the quantities used in the matrix element calculation 
include the rejection of events with either an additional jet of $\Et > 12$  GeV 
or a second high-$\Pt$ charged lepton. The latter reduces $Z$+jets,
$t\bar{t}$, and leptonic diboson backgrounds. 
For events with an electron candidate, there is a significant 
background from production of multiple jets (multi-jet in the following) by quantum chromodynamical (QCD) processes, where the electron is faked by a hadronic jet.  The ME method deals with this background by 
applying stringent selection criteria, while the $M_{jj}$ method assigns a systematic uncertainty to the background shape.
%which is more problematic for the ME method. 
The reduction of the multi-jet QCD background in the ME analysis is achieved by raising the $\MET$ cut to 40~GeV, requiring 
% the transverse mass of the leptonic $W$ 
$M_{T}(W)>70$~GeV/$c^2$, and imposing additional cuts on the angles between the jets, the lepton, and the $\met$~\cite{pdong}. There is a less
stringent requirement of $M_{T}(W)>10$~GeV/$c^2$ imposed on muon events
to reduce the QCD background in that channel.

%%%%%%%%%%%%%%%%%%%%%%%%%%%%%%%

%\section{Modeling and backgrounds}
After these selections for both methods, 
the dominant background to the diboson signal is a $W$ boson produced with accompanying jets ($W$+jets),
where the $W$ decays leptonically.  Smaller but non-negligible
backgrounds come from QCD multi-jet (where one jet mimics a lepton
signature), $Z$+jets, $t\bar{t}$, and single
top production.  QCD multi-jet events are modeled using data with loosened 
lepton selection criteria. 
Signal and other background processes are modeled using event
generators and a {\sc geant}-based CDF II detector simulation.  The diboson
signals and the $t\bar{t}$ and single top backgrounds are
simulated using the {\sc pythia} event generator~\cite{pythia}.  The
$W$+jets and $Z$+jets backgrounds are simulated using the tree-level
event generator {\sc alpgen}~\cite{alpgen}, with an interface to 
{\sc pythia} providing parton showering and hadronization.

The normalization of the $Z$+jets background is based on the measured cross 
section while for $t\bar{t}$  and single top backgrounds the NLO predicted 
cross section is used~\cite{xsections}.  The efficiencies for the $Z$+jets, $t\bar{t}$, 
and single top backgrounds are estimated from simulation.
The normalization of the QCD background is
estimated by fitting the $\MET$ spectrum in data to the sum of all
contributing processes, where the QCD and $W$+jets normalizations
float in the fit.  
%The normalization of the $W$+jets background is a
%free parameter in the final signal extractions from both methods.  
In the final signal extractions from both methods, the multi-jet
QCD background is Gaussian constrained to the result of this $\MET$ fit
and the $W$+jets background is left unconstrained. 
%%%% make sure that the above is indeed true for both analyses 

%%%%%%%%%%%%%%%%%%%%%%%%% Mjj
We now describe the methodology and results from 
each technique. In the $M_{jj}$ method we extract the signal
fraction from the data by performing a $\chi^2$ fit to the 
dijet invariant mass spectrum separately for electron and
muon events. Templates of $M_{jj}$ distributions are constructed
with the multi-jet QCD background, the signal $WW+WZ$ processes, and the sum of the electroweak backgrounds ($Z+{\rm jets}$,
$W + {\rm jets}$, and $t\bar{t}$ production).

%The normalizations of the electroweak background and the signal are unconstrained 
%parameters in the fit, while the
%multi-jet QCD background is Gaussian constrained to the result of the \MET~fit
%described above, with an uncertainty of 20\%.
%The overall normalization of the sum of signal and
%all backgrounds is also a free parameter in the signal extraction fit.

\begin{figure}[!htb]
\includegraphics[width=1\columnwidth]{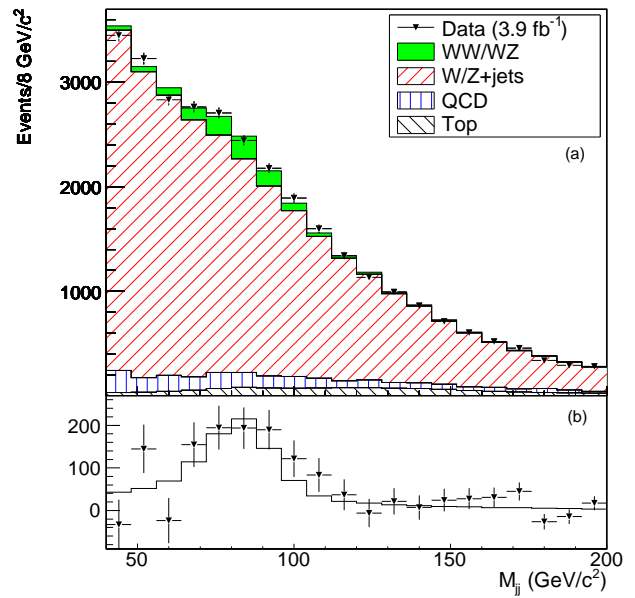}
\caption{\label{fig:projections} Dijet invariant mass distribution
of reconstructed $W/Z \rightarrow jj$ candidates compared to the
fitted signal and background components (a), and for the
corresponding background subtracted distribution (b).
}
\end{figure}

Figure~\ref{fig:projections} shows the fit results superimposed on data after the electron and muon samples are combined.
%shows the projection of the  superimposed on data 
%after the electron and muon samples are combined. 
Also shown is the data $M_{jj}$ distribution after having subtracted the estimated background, superimposed on the signal Monte Carlo normalized to the fit result.
%The $\chi^2$ fits return $428 \pm 177 {\rm (stat)} \pm 56 {\rm (syst)}$ signal events 
%in the electron sample and $650 \pm 149 {\rm (stat)} \pm 66 {\rm (syst)}$
%signal events in the muon sample. Combining the two channels, 
%we estimate a total of $1079 \pm 232 {\rm (stat)} \pm 86 {\rm (syst)}$ 
%$WW/WZ \rightarrow \ell \nu jj$ events, 
%corresponding to an observed significance of 4.6 $\sigma$ where 4.9 $\sigma$ is expected.
Combining the two $\chi^2$ fit results we get a total of
$1079 \pm 232 {\rm (stat)} \pm 86 {\rm (syst)}$ 
$WW/WZ \rightarrow \ell \nu jj$ events, of which
about 60\% are muon events and 40\% are electron events. 
%%%
% The significance of this result is evaluated as 
% ${\rm \frac{N_{signal}(measured)}{N_{signal}(observed)}/\sqrt{\sigma_{stat}^2 + \sigma_{syst}^2}}$
% where $\sigma_{stat}$ and $\sigma_{syst}$ are the statistical and systematic uncertainties
% on the ratio ${\rm \frac{N_{signal}(measured)}{N_{signal}(observed)}}$.
%%%
The observed significance is 4.6$\sigma$ where 4.9$\sigma$ is expected.
%which is obtained by combining the separate results from the electron and muon channels.
The resultant  
$WW + WZ$ production cross section measurement is  
$\sigma_{WW+WZ} = 14.4 \pm 3.1{\rm (stat)}\pm 2.2 {\rm (syst)}\ {\rm pb}$. The sources
of systematic uncertainty in this measurement are discussed together with those
from the ME method below.

%

%%%%%%%%%%%%%%%%%%%%%%%%%%%% ME

\begin{table}[ht]
\caption{\label{tab:yields}Expected and observed event yields after the ME method selection in
$2.7\,{\rm fb^{-1}}$ of data.}
\begin{ruledtabular}
\begin{tabular}{lc}
Process & Predicted event yield \\
\hline
$WW$ signal & $446 \pm 29$ \\
$WZ$ signal & $79 \pm 6$ \\
$W$+jets & $10175 \pm 305$ \\
$Z$+jets & $584 \pm 88$ \\
QCD multi-jet & $283 \pm 113$ \\
$t\bar{t}$ + single top & $241 \pm 29$ \\
\hline
Observed & 11812 \\
\end{tabular}
\end{ruledtabular}
\end{table}

%\section{Matrix element method} 
In the ME method a
probability density $P(x)$ that an event was produced by a given
process is determined using the standard model differential cross
section for that process.  For an event with measured quantities $x$,
we integrate the appropriate differential cross section $d\sigma(y)$
over the partonic quantities $y$ convolved with the parton
distribution functions (PDFs) and a transfer function:
\begin{equation}
\label{eqn:evtprob}
P(x)=\frac{1}{\sigma}\int d\sigma(y)dq_1dq_2f(q_1)f(q_2)W(y,x).
\end{equation}
The PDFs ($f(q_1)$ and $f(q_2)$) are evaluated according to the CTEQ5L parameterization~\cite{CTEQ}.  The transfer function $W(x,y)$ relates $x$ to $y$, encoding the effects of the detector resolution.  The momenta of electrons, muons, and the angles of jets are assumed to be measured exactly and a mapping of measured jet energy to partonic energy is derived using the full detector simulation.  The integration is performed over the
energy of the partons and the longitudinal momentum of the neutrino.  The matrix element is
calculated with tree-level diagrams from {\sc madgraph}~\cite{madgraph}.  
Event probability densities are calculated for the signal processes as well as for $W$+jets and single top
background processes.  The event probabilities are
combined into an event probability discriminant:
$EPD=P_{signal}/(P_{signal}+P_{background})$, where $P_{signal} =
P_{WW}+P_{WZ}$ and $P_{background} = P_{W+jets}+P_{single\; top}$.  We
make templates of the $EPD$ for all signal and background processes
and ultimately extract the signal using a fit of the observed $EPD$
distribution to a sum of the templates.  The
expected event yields are as shown in Table~\ref{tab:yields} for the
ME method's event selection.

\begin{figure}[ht]
\includegraphics[width=0.96\columnwidth]{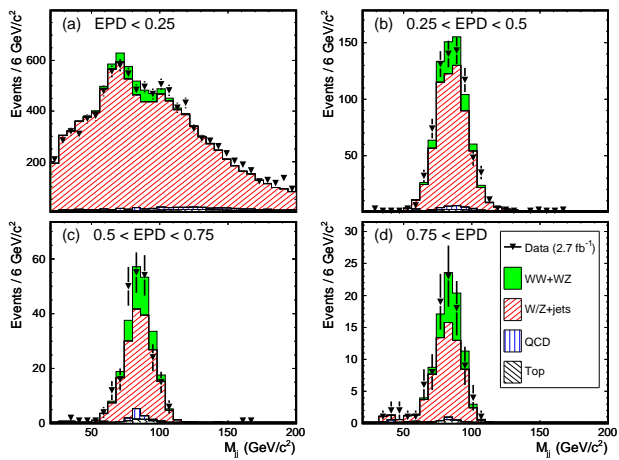}
\caption{$M_{jj}$ for events with (a) $EPD < 0.25$, (b) $0.25 < EPD < 0.5$,
(c) $0.5 < EPD < 0.75$, and (d) $EPD > 0.75$.}
\label{fig:EPDbins}
\end{figure}

Figure~\ref{fig:EPDbins} shows the dijet mass in bins of $EPD$.  Most of
the background events have low $EPD$.  Events with $EPD > 0.25$ have a dijet mass peak close to the expected $W/Z$ resonance, and the signal-to-background ratio improves
with increasing $EPD$.

Before comparing the observed $EPD$ to the prediction, we validate the
Monte Carlo modeling of the quantities that enter the matrix element
calculation.  We compare the observed distributions to the predicted
ones in control regions with very little signal and also in the
signal-rich region.  The different regions are chosen according to the
invariant mass of the two-jet system ($M_{jj}$): the signal-rich
region has $55<M_{jj}<120$ GeV and the control regions cover the
rest of the $M_{jj}$ range.  We also check the modeling of the
properties (mass, \Pt,  and $\eta$) of the leptonic $W$ boson and the
hadronic $W$ or $Z$ boson candidate.  All of these quantities are well described
by the simulation for our event selection.  There is a small
discrepancy in the description of $M_{jj}$ in the control regions, as
is visible in the low-$EPD$ region of Figure~\ref{fig:EPDbins}.
Associated with this discrepancy we assign a systematic mismodeling
uncertainty which is derived in the control regions and extrapolated
through the signal region.  This uncertainty has a negligible effect
on the results, because most background events lie in the first few
bins of the $EPD$ distribution.  Small changes in modeling of those
background events do not change the shape of the $EPD$.

\begin{figure}[ht]
\includegraphics[scale=0.5,width=1.0\columnwidth]{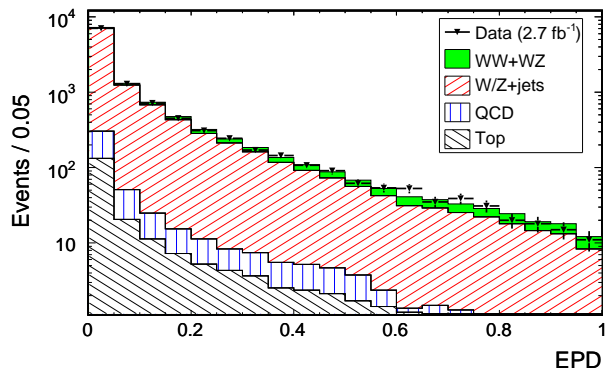}
\caption{Observed $EPD$ distribution superimposed on distribution expected from simulated processes.}
\label{fig:EPD}
\end{figure}

The observed and predicted $EPD$s are shown in Figure~\ref{fig:EPD}.
We use a binned-likelihood fit of the observed $EPD$ to a sum of
templates, testing both a background-only hypothesis and a
signal-plus-background ($s+b$) hypothesis.  Systematic uncertainties,
discussed further below, are included in the fit as constrained
parameters.  We perform pseudo experiments to calculate the
probability ($p$-value) that the background-only discriminant
fluctuates up to the observed result (observed $p$-value) and up to
the median expected $s+b$ result (expected $p$-value).  We observe a
$p$-value of $2.1 \times 10^{-7}$, corresponding to a signal
significance of 5.4$\sigma$, where $5.1\sigma$ is expected.  The
observed $WW+WZ$ cross section is $\sigma_{WW+WZ} = 17.7 \pm 3.1{\rm
(stat)} \pm 2.4 {\rm (syst)}\ {\rm pb}$.

%%%%%%%%%%%%% systematics -- need to talk about whether this is the best place for it

We consider several sources of systematic uncertainty in
both methods, taking
into account their effect on both the signal acceptance and the shape
of the background and signal templates.  The uncertainty on the
normalization of the backgrounds is taken as part of the statistical
uncertainty.  
%In the $M_{jj}$ method the largest systematic uncertainties
%are due to the modeling of the electroweak and QCD shapes (about 8\% and
%6\% respectively) and the uncertainty of the jet energy scale (6\%).
%In the ME method the jet energy scale is the largest systematic
%uncertainty. Its effect is quantified by varying the jet energies in
%the simulated samples by $\pm 1 \sigma$.  We assign both an acceptance
%uncertainty and a shape uncertainty on the signal templates due to this
%effect resulting in a total jet energy scale systematic of about 10\%.
In the $M_{jj}$ method the largest systematic uncertainties
are due to the modeling of the electroweak and QCD shapes, about 8\% and
6\% respectively. In the ME method the uncertainty in the jet energy scale 
is the largest systematic uncertainty, at about 10\%, which includes 
contributions both from the signal acceptance and from the shapes of the 
signal templates. In the $M_{jj}$ method this uncertainty is about 6\%.
%%%%
Both methods include an  uncertainty of about 5\% due to initial and 
final state radiation and a 6\% uncertainty on the integrated luminosity. 
Smaller contributions arise from PDFs, jet energy 
resolution, the factorization and renormalization scales used 
in the $W$+jets simulation, and trigger and lepton identification efficiencies.

%%%%%%%%%%%%%% correlations %%%%%%%%%%%%%%%%%%
One measure of how the two methods are correlated is the 
expected overlap of $WW+WZ$ signal.
Accounting for the different integrated luminosities used, 15\% of
the signal in the $M_{jj}$ analysis is common to that in the $EPD$ analysis.
Conversely, 29\% of the signal in the $EPD$ analysis is common to that in
the $M_{jj}$ analysis. This corresponds to a statistical correlation 
of about 21\%. If we assume the systematic uncertainties are 100\% correlated, then
the total correlation between the two analyses is 49\%, leading to a
combined~\cite{lyons1988} result of $\sigma_{WW+WZ} = 16.0 \pm 3.3{\rm (stat + syst)}\ {\rm pb}$.
%The central value of this combined result does not depend on what
%assumption we make for the correlation between systematic uncertainties, 
%and only changes the total error by a few percent for any reasonable assumption.
Because the total uncertainties on the two input measurements are 
similar, the combined central value does not depend significantly on
the correlation assumed. The total uncertainty in the combined result
increases with increasing correlation and we quote the value assuming
maximum possible correlation.
%In addition, it is interesting to look at the correlations between
%these analysis and the CDF $WW + WZ + ZZ$ observation in the $\met + jets$
%analysis~\cite{cdfmetjj}, which has a larger $\met$ requirement but
%also includes events with leptons. About 15\% of the $WW + WZ$ signal
%from this analysis appears in the analyses presented here.
The signal overlap with the CDF $WW + WZ + ZZ$ observation in the $\met$+jets
channel~\cite{cdfmetjj} is also studied. While that analysis requires
much larger $\met$, it does not veto events with identified leptons.
We found that about 15\% of the  $WW + WZ$ signal from the $\met$+jets
analysis appears in the analyses presented here.

In summary, we observe $WW+WZ$ production in the lepton plus 
jets plus $\MET$ final state.  We perform two searches: 
one seeking a resonance on top of a smoothly falling dijet mass distribution,
and another building a discriminant using a matrix element technique. 
The combined $WW + WZ$ cross section from these two methods is measured
to be  $\sigma_{WW+WZ} = 16.0 \pm 3.3{\rm (stat + syst)}\ {\rm pb}$,
%The $WW+WZ$ cross section is measured to be
% $14.4 \pm 3.1 {\rm (stat.)} \pm 2.2 {\rm (syst.)}\ {\rm pb}$ and 
% $17.7 \pm 3.1 {\rm (stat.)} \pm 2.4 {\rm (syst.)}\ {\rm pb}$
%with the $M_{jj}$ and matrix-element searches
%respectively. 
in good agreement with the prediction of
$16.1\pm0.9$~pb~\cite{VVtheory}.  Measurements of these diboson
processes are tests of electroweak theory and a necessary step
toward validating Higgs boson search techniques at the Tevatron.

\begin{acknowledgments}
We thank the Fermilab staff and the technical staffs of the
participating institutions for their vital contributions. This work
was supported by the U.S. Department of Energy and National Science
Foundation; the Italian Istituto Nazionale di Fisica Nucleare; the
Ministry of Education, Culture, Sports, Science and Technology of
Japan; the Natural Sciences and Engineering Research Council of
Canada; the National Science Council of the Republic of China; the
Swiss National Science Foundation; the A.P. Sloan Foundation; the
Bundesministerium f\"ur Bildung und Forschung, Germany; the World
Class University Program, the National Research Foundation of Korea;
the Science and Technology Facilities Council and the Royal Society,
UK; the Institut National de Physique Nucleaire et Physique des
Particules/CNRS; the Russian Foundation for Basic Research; the
Ministerio de Ciencia e Innovaci\'{o}n, and Programa
Consolider-Ingenio 2010, Spain; the Slovak R\&D Agency; and the
Academy of Finland.
\end{acknowledgments}


\begin{thebibliography}{99}
\bibitem{coord} We use a cylindrical coordinate system with its origin in the center 
of the detector, where $\theta$ and $\phi$ are the polar and azimuthal angles, respectively,
and pseudorapidity is $\eta= - \rm ln$ $\rm tan(\theta/2)$. 
The transverse momentum of a charged particle is $p_T = p\sin\theta$, where $p$ is the momentum of the charged particle track.  The transverse energy, $E_T = E\sin\theta$, is measured using the calorimeter.
The missing $E_T$ ($\METVEC$) is defined by $\METVEC = -\sum_iE_T^i{\hat n}_i$,
where ${\hat n}_i$ is a unit vector perpendicular to the beam axis and pointing at the
$i^{th}$ calorimeter tower.  $\METVEC$ is corrected for high-energy muons and jet energy corrections.
We define $\MET=|\METVEC|$. The transverse mass of the $W$ is defined as $M_T (W)=\sqrt{2p_T^l \MET (1-\cos(\Delta \phi^{l\nu}))}$.
\end{thebibliography}
\end{document}